\DocumentMetadata{}
\documentclass[acmsmall]{acmart}

\usepackage{listings}

\usepackage{xcolor}
\usepackage{soul}

\setcopyright{none}
\usepackage{caption}
\definecolor{codegreen}{rgb}{0,0.6,0}
\definecolor{codegray}{rgb}{0.5,0.5,0.5}
\definecolor{codepurple}{rgb}{0.58,0,0.82}
\definecolor{backcolour}{rgb}{0.95,0.95,0.92}
\usepackage{array}
\lstdefinestyle{mystyle}{
    commentstyle=\color{codegreen},
    keywordstyle=\color{magenta},
    numberstyle=\tiny\color{codegray},
    stringstyle=\color{codepurple},
    basicstyle=\ttfamily\footnotesize,
    breakatwhitespace=false,         
    breaklines=true,                 
    captionpos=b,                    
    keepspaces=true,                 
    numbers=left,                    
    numbersep=5pt,                  
    showspaces=false,                
    showstringspaces=false,
    showtabs=false,                  
    tabsize=2,
    frame=single,
    aboveskip=15pt,
    belowskip=15pt
}

\AtBeginDocument{%
  \providecommand\BibTeX{{%
    \normalfont B\kern-0.5em{\scshape i\kern-0.25em b}\kern-0.8em\TeX}}}

\copyrightyear{2024}
\acmYear{2024}
\setcopyright{none}
\acmConference[PEARC '24]{Practice and Experience in Advanced Research Computing}{July 21--25, 2024}{Providence, RI, USA}
\acmBooktitle{Practice and Experience in Advanced Research Computing (PEARC '24), July 21--25, 2024, Providence, RI, USA}
\acmDOI{10.1145/3626203.3670557}
\acmISBN{979-8-4007-0419-2/24/07}

\begin{document}
\title[Open Science Data Federation]{Open Science Data Federation - operation and monitoring}

\author{Fabio Andrijauskas}
\email{fandrijauskas@ucsd.edu}
\orcid{0000-0002-1254-8570}
\affiliation{%
  \institution{University of California - San Diego}
  \streetaddress{9500 Gilman Dr, La Jolla}
  \city{San Diego}
  \state{CA}
  \country{US}
  \postcode{92093}
}

\author{Derek Weitzel}
\email{dweitzel@unl.edu}
\orcid{0000-0002-8115-7573}
\affiliation{%
  \institution{University of Nebraska-Lincoln}
  \streetaddress{1400 R St}
  \city{Lincoln}
  \state{NE}
  \country{US}
  \postcode{68588}
}

\author{Frank K W\"urthwein}
\email{fkw@physics.ucsd.edu}
\orcid{0000-0001-5912-6124}
\affiliation{%
  \institution{University of California - San Diego}
  \streetaddress{9500 Gilman Dr, La Jolla}
  \city{San Diego}
  \state{CA}
  \country{US}
  \postcode{92093}
}
\renewcommand{\shortauthors}{Andrijauskas et al.}

\begin{abstract}
Extensive data processing is becoming commonplace in many fields of science. Distributing data to processing sites and providing methods to share the data with collaborators efficiently has become essential. The Open Science Data Federation (OSDF) builds upon the successful StashCache project to create a global data access network. The OSDF expands the StashCache project to add new data origins and caches, access methods, monitoring, and accounting mechanisms. Additionally, the OSDF has become an integral part of the U.S. national cyberinfrastructure landscape due to the sharing requirements of recent NSF solicitations, which the OSDF is uniquely positioned to enable. The OSDF continues to be utilized by many research collaborations and individual users, which pull the data to many research infrastructures and projects. 
\end{abstract}



\maketitle

\section{Introduction}

As the scale of data and science complexity increases, research must look for additional resources to manage storage and compute requirements in memory or processor speed \cite{schultz2023icecube,COLUCI2023171}. Regardless of whether projects need large computational throughput, e.g. Open Science Consortium (OSG)  \cite{10.1145/3332186.3332212} or computational performance, e.g. ACCESS \cite{accessci}, or the National Research Platform \cite{nationalresearchplatform} an unquestionable challenge is pervasive across all forms of computational access: what is the most efficient way to deliver data to a compute host when there are always limits on local infrastructure. For many collaborative research projects, the answer is to place input data wherever there is storage capacity, maybe at the site of an experiment and then find ways to deliver that data to a compute site as needed by a computational pipeline. For a domain researcher, it is a challenge to bridge the geographical distance separating their data from a CPU that will process them \cite{weitzel2017data}. The Open Science Data Federation (OSDF) is a proven data access framework that provides the infrastructure and the tools to access data globally, implementing the notion of "Any Data, Anytime, Anywhere". When we read our favorite newspaper online, we don't ask for it to first be delivered to our laptops, or phones. OSDF does the same for data access for computation. It replaces a focus on data transfer with a focus on data access by the application. Figure 1 shows the current global deployments of OSDF caches and origins. 

\begin{figure}[ht]
         \centering
         \includegraphics[width=\textwidth]{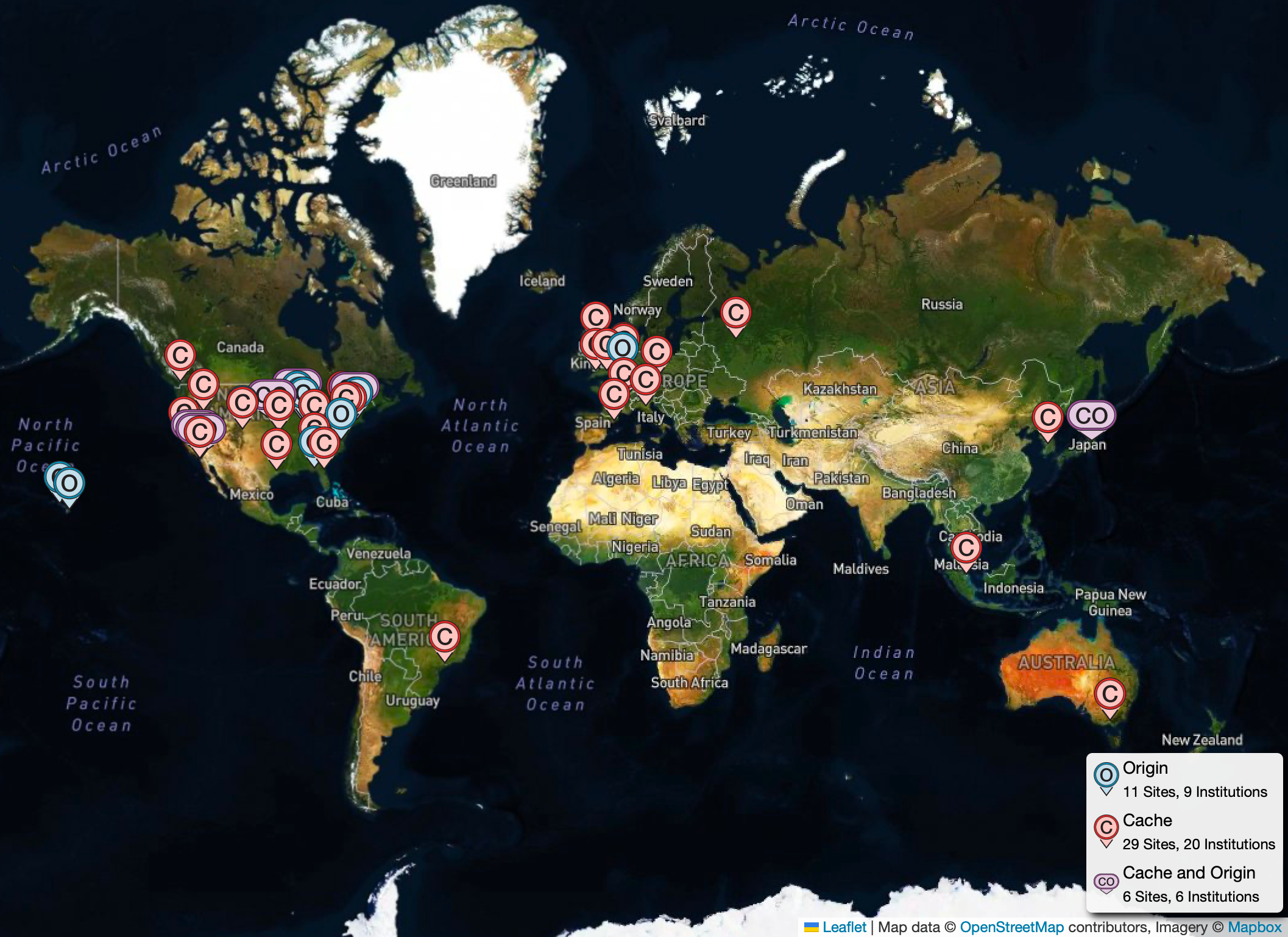}
         \caption{OSDF caches and origins worldwide.}
         \label{osdfn1}
\end{figure}

OSDF is the evolution of an OSG project, StashCache \cite{10.1145/3332186.3332212,Fajardo_2018}. Together, the OSG and OSDF create a unique research environment, providing computational throughput capacity and efficient access to their data for users \cite{schultz2023icecube}. To meet demand and growth, the OSDF ecosystem needs to improve in tools and capacity continuously. It also needs to ensure that data are reliably delivered at compute sites and staff are empowered to troubleshoot potential shortfalls as early as possible. 
In this work, we show improvements in accounting, available storage space for caches and origins, and enhancements in monitoring the quality of the service delivery.

\section{Background}

\begin{figure}[ht]
         \centering
         \includegraphics[scale=.3]{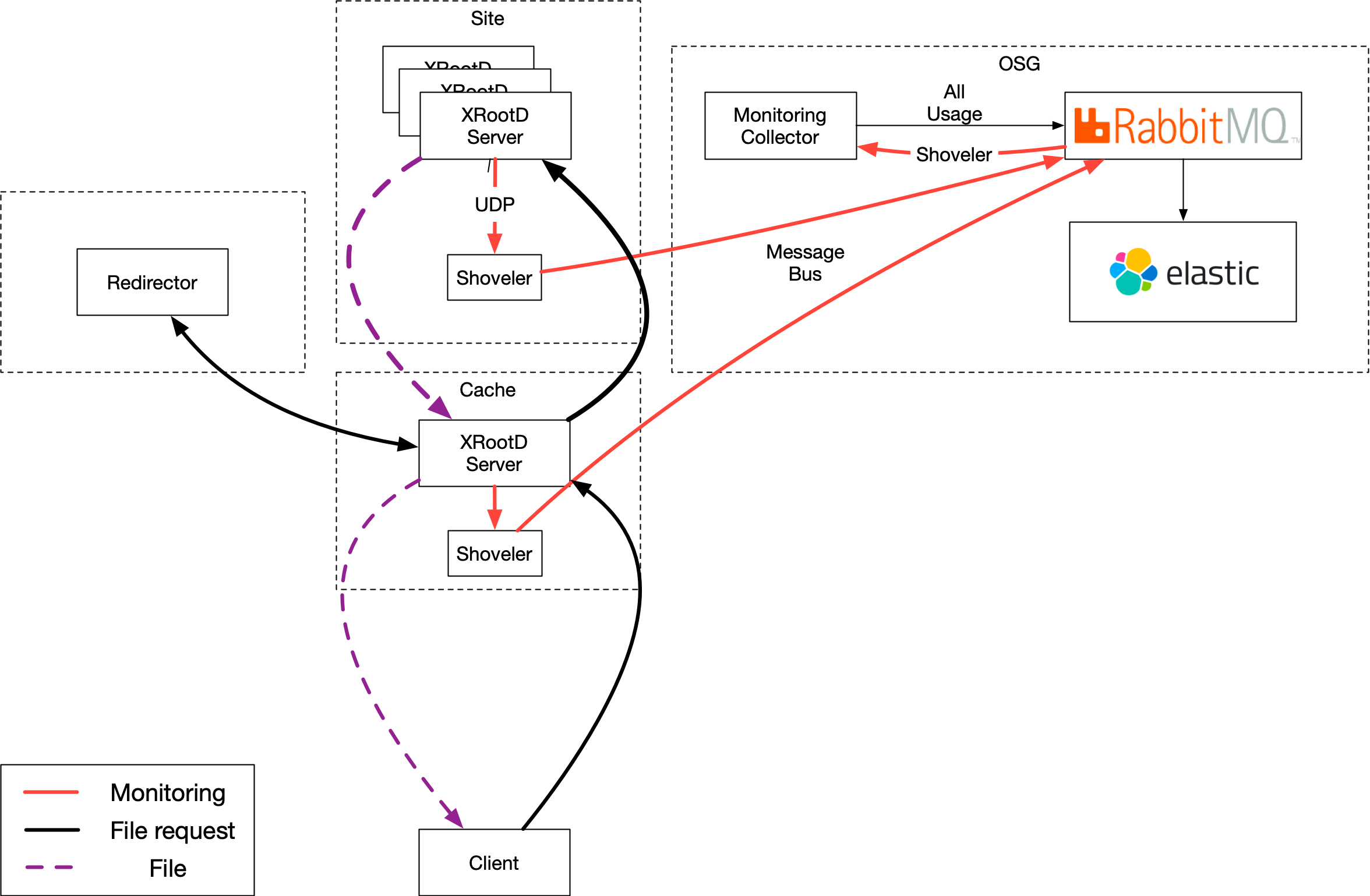}
         \caption{The Open Science Data Federation uses XrootD, and several other tools to serve data to execution points via client requests.}
         \label{osg}
   \end{figure}

In order to serve data to computational workflows running on the distributed computing infrastructure, the OSG utilizes the Open Science Data Federation \cite{10.1145/3332186.3332212}. At the crux of this data delivery framework are the concepts of "origin," "caches," and "redirectors," all implemented as services via the XrootD \cite{xrootd-paper} software framework, which allows for low latency and scalable data access. Figure~\ref{osg} highlights the integration of these components. Origins effectively refer to the backend storage hosting project data. In the context of OSDF, an origin refers to the XRootD configuration that allows access to the storage via a data transfer node that mounts a project directory. Multiple origins are tied into a tree structure that connects to a redirector which communicates with the cache network. Applications generally access the OSDF via the closest cache to a computing site. The determination of the closest cache is via GeoIP \cite{10.1145/3332186.3332212}. The most efficient way to deploy and manage OSDF software and services (at origins or caches) is with containers on a federated Kubernetes infrastructure. The federated model allows operators to monitor data access and debug issues effectively. 

Figure \ref{osg} shows the primary file access. The black line represents a file request by a job, the purple dashed lines represent file access, and the orange shows the monitoring process. When a job requests a file, it queries the file from the nearby cache. If the file is on the cache, the job receives the file. However, if the file is unavailable on the cache, the cache queries the redirector of its location at an origin. Once the file location is determined, the cache requests the file from the origin and makes it available to the job. XRootD provides operators with the ability to get informational streams across all levels of interaction. We will discuss this briefly in Section~\ref{monitoring}.
Data access on OSDF is enabled via multiple ways, with the majority being requested by jobs running in pools managed by the Open Science Consortium (OSG). The network of Caches and Origins can serve data for multiple projects and science collaborations to points of execution and facilitate the acceleration of scientific discovery. This capability is being promoted by the U.S. National Science Foundation (NFS) via the Campus Cyberinfrastructure (CC*) program that awards funds for storage infrastructure that will host OSDF origins and caches. 
\url{https://opensciencegrid.org/campus-cyberinfrastructure.html} and \url{https://beta.nsf.gov/funding/opportunities/campus-cyberinfrastructure-cc}. Most OSDF hosts are deployed using NRP and Kubernetes using a GitHub DevOps architecture.

\section{New caches and origins}

Funding accessibility, along with the success of past deployments, help pave the way for deploying more federated caches and origins to geographically span all of the U.S. We provide below a list of recent deployments:

\begin{enumerate}
    \item 1 cache (50 TB) and one origin (1.6TB): San Diego Supercomputer Center.
    \item 1 cache (42 TB) and one origin (1.2 PB):  University of Nebraska-Lincoln.
    \item 1 cache (29 TB) and one origin (1.2PB): Massachusetts Green High-Performance Computing Center.
    \item New caches on Internet2 hubs: Boise - Idaho (42 TB), Houston - Texas (in progress), Jacksonville - Florida (42 TB), Denver  - Colorado (42 TB), and Northeastern University Boston. Existing Internet2 caches in Chicago, New York, and Kansas City.
    \item 1 cache (7 TB) and one origin (7 TB): The University of Tokyo. 
    \item 1 cache (14 TB): The National Center for Atmospheric Research (NCAR), Boulder.
    \item 1 origin (337 TB): University of Nebraska-Lincoln.
\end{enumerate}

The location of OSDF infrastructure is primarily determined by proximity to a University campus, an institution, or an Internet2 point of presence (PoP). Furthermore, the institution must have an affiliation to a project or a collaboration. As shown in Figure~\ref{osg}, the monitoring streams from OSDF are collected in an Elasticsearch database, which yields an abundance of information that we can mine for performance and utilization metrics of the service \cite{Deng_2023}.

\begin{figure}[ht]
         \centering
         \includegraphics[scale=.15]{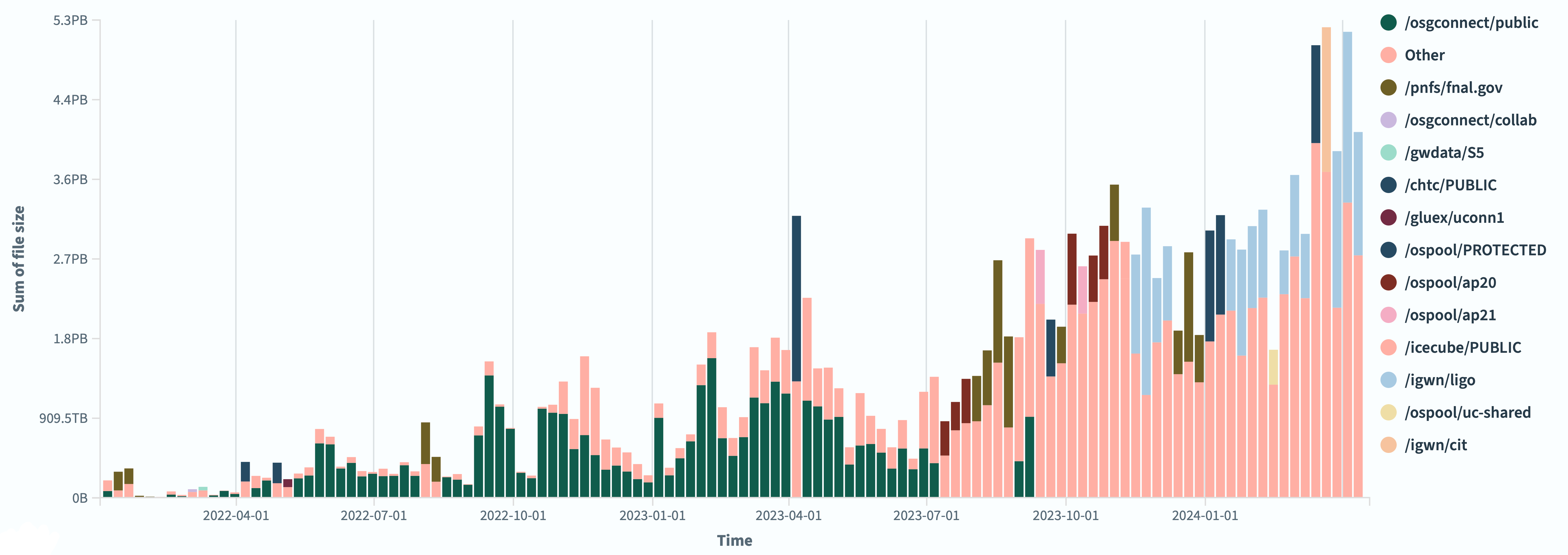}
         \caption{The monthly evolution of cache storage per project namespace shows the steady growth of the OSDF utilization by research projects. }
         \label{osdfreq}
   \end{figure}

We show in Figure \ref{osdfreq} the aggregated storage utilization values of the caches per month of top project namespaces; there are 375 project namespaces in OSDF at present. Table 1 showcases some of the top level statistics for the OSDF infrastructure during the past year, Table 2 shows OSDF top-level statistics for 2022-2023.

\begin{table}
\scriptsize

\begin{minipage}[b]{.40\textwidth}
  \centering
  \begin{tabular}{ | l | l |}
    \hline
Metric & Value \\ [0.2ex] 
 \hline\hline
Caches & 37 across 14 Inst.\\ 
 \hline
Origins & 15 across 6 Inst. \\
 \hline
Transfers & 5,861,721,390 \\
 \hline
Total size of transfers & 294.7PB\\
\hline
Number of checks/monitoring & 352 \\
\hline
Number of Kubernetes Pods & 82 \\
\hline

  \end{tabular}
    \label{generalstats}
\scriptsize  \caption{\scriptsize OSDF top-level statistics, 2022-2023.}

\end{minipage}\qquad
\begin{minipage}[b]{.40\textwidth}
  \centering
  \begin{tabular}{ | l | l |}
\hline
 Project & Files request \\ [0.5ex] 
 \hline\hline
 LIGO & 6893702577 \\ 
 \hline
 LIGO user-specific & 565677512 \\ [1ex] 
 \hline
  fnal.gov - Nova & 492252998 \\ 
   \hline
  fnal.gov - Minerva & 435978085 \\ 
     \hline
  fnal.gov - Dune & 412684786 \\
      \hline
  OSG Collaborations  & 226827338 \\
        \hline
  fnal.gov - uboone  & 156274029 \\
          \hline
 LIGO gwdata - O3a - uboone  & 145505762 \\
            \hline
  LIGO gwdata - O2 - uboone  & 103851076 \\
              \hline
  OSG user & 98685798 \\
                \hline
  Other & 766798555 \\
 \hline
  \end{tabular}
\scriptsize  \caption{\scriptsize Files requested by projects between 2018 and 2023.}
  \label{generalstats2}
\end{minipage}
\end{table}

\begin{itemize}
    \item LIGO - Laser Interferometer Gravitational-Wave Observatory \url{https://www.ligo.caltech.edu}
    \item NOvA - NuMI Off-axis ve Appearance - \url{https://novaexperiment.fnal.gov}
    \item MINERvA -  (Main Injector Neutrino ExpeRiment to study v-A interactions) - \url{https://minerva.fnal.gov}
    \item DUNE - Deep Underground Neutrino Experiment - \url{https://lbnf-dune.fnal.gov}
    \item MicroBooNE  - large 170-ton liquid-argon time projection chamber - \url{https://microboone.fnal.gov}
\end{itemize}

\section{End-to-end Monitoring of OSDF}\label{monitoring}

As it is critical that we ensure the operational stability and availability of the OSDF infrastructure, we employ extensive monitoring of all the components and perform regular tests that check the status of the end-to-end service delivery. Figure~\ref{osg} provides a graphical description of the service flow. As mentioned previously, when a client requests a file from a nearby cache and it is not there, then the cache queries the file location via the redirector and requests it from the origin. This flow of queries and decision-making is captured by an XRootD stream, one of several supported in XRootD. For example, the f-stream captures information about the file access. The g-stream has information about the cache hit and misses. All of these collections of messages from the cache or the origin are sent to a Shoveler service. The Shoveler is a tool that collects the UDP XrootD streams and sends them to the RabbitMQ by TCP. After RabbitMQ receives the message, a monitoring collector logs the message. The queue of messages in RabbitMQ is then indexed in the Elasticsearch database, from where statistics and useful information can be mined.

In addition to the XRootD monitoring streams, we employ tests in Checkmk (\url{https://checkmk.com}) designed to check the end-to-end OSDF file access. These aim to detect issues with the infrastructure before they become problematic for users. For example, we would be interested for a view into the state of the network bandwidth between origins and caches. A daily cadence test checking transfer rates between origins to caches with sample data can reveal problems, allowing time for teams to respond before a production pipeline running at execution points starts suffering from performance degradation in the data delivery. We list below a comprehensive list of tests we perform on the OSDF infrastructure that ensures that staff get advanced warnings on the state of the service: 

\begin{enumerate}
    \item Monitor authenticated access using certificates and tokens (caches).
    \item Check the protected files on OSDF. OSDF has a structure of protected files using SciTokens \cite{10.1145/3219104.3219135} or certificates. Checking if accessing the file without the correct credentials is possible (caches).
    \item Monitor the number of messages the shoveler processed (caches and origins).
    \item Check the size of the shoveler queue (caches).
    \item Check if copying files from public and private origins (caches and origins) is possible.
    \item Verify the transfer rate considering the host location (caches and origins).
    \item Verify the SSL certificate (caches and origins).
    \item Check the load of each cache and origin in NRP (caches and origins).
    \item Check the CVFMS access (caches).
    \item Verify if the redirector is running and operational. 
\end{enumerate}

\section{Conclusion}
We have presented a list of operational and monitoring upgrades of the Open Science Data Federation. Storage space tracking, monitoring checks, monitoring stream collections, and adding new caches has increased the robustness and capacity of this data delivery infrastructure. New monitoring processes provide a way to detect problems before the users. New caches expand nearby proximity access to compute sites. New origins increase storage capacity. File access streams allow for more efficient debugging and support. These improvements create a way of the data delivery experience in the computing ecosystem fostered by the Open Science Consortium.

New software and caching techniques are being added frequently to the Open Science Data Federation software.  In the future, we will incorporate software from the Pelican \cite{pelicanproject} project, which will increase visibility for cache and origin owners, as well as provide better traceability of usage by clients.

\begin{acks}
This work was supported in part by National Science Foundation (NSF) awards \#1836650, CNS-1730158, ACI-1540112, ACI-1541349, OAC-1826967, OAC-2030508, OAC-2112167, CNS-2100237, CNS-2120019, PHY-2323298, the University of California Office of the President, and the University of California San Diego's California Institute for Telecommunications and Information Technology/Qualcomm Institute. Thanks to CENIC for the 100Gbps networks. Also, Fabio would like to acknowledge Erick Baltes and Paschalis Paschos.
\end{acks}

\bibliographystyle{ACM-Reference-Format}
\bibliography{bibliography}

\end{document}